\documentclass[aip, pop, twocolumn, reprint, groupedaddress]{revtex4-1}
\usepackage{graphicx}

\begin{document}

\title{Observation of ionization-mediated transition from collisionless 
interpenetration to collisional stagnation during merging of two supersonic 
plasmas}

\author{Auna L. Moser}\thanks{Now at General Atomics, San Diego, CA; electronic mail:  mosera@fusion.gat.com.}
\author{Scott C. Hsu}\email{scotthsu@lanl.gov.}
\affiliation{Physics Division, Los Alamos National Laboratory, Los Alamos, New  
Mexico 87545}

\date{\today}

\begin{abstract}    
We present space- and time-resolved experimental data of head-on-merging, 
supersonic plasma jets (of an argon/impurity mixture) in an initially 
collisionless regime for counter-streaming ions.  The merging begins with
collisionless interpenetration followed by a transition to collisional 
stagnation.  The transition occurs due to an experimentally inferred rising mean-ionization 
level, which rapidly reduces the counter-streaming ion--ion mean free path.
The measurements demonstrate a specific mechanism by 
which a collisionless interaction transitions to a collisional one and 
constrain collisionality and ionization models for plasmas with complex equation of state.
 
\end{abstract}


\maketitle


The dynamics of colliding plasmas plays an important role in, e.g., hohlraum 
plasmas in inertial confinement fusion,\cite{atzeni04} astrophysical shock 
waves,\cite{sagdeev91sa} and applications such as pulsed laser deposition.
\cite{luna07jap} Colliding plasma interactions can often be in a regime that 
is neither purely collisional nor purely collisionless, a complicated situation 
for modeling the interactions.
\cite{larroche93pfb,rambo95pop,jones96pop,thoma13pop} 
Previous experiments 
\cite{bosch92pfb,rancu95prl,wan97pre,ross12pop,ross13prl,swadling13pop_b,swadling14prl,al-shboul14pop} have 
reported observations of plasma interpenetration on millimeter scales in laser-driven
or wire-array Z-pinch experiments.
Here we present 
experimental results from the head-on collision of two larger-scale, railgun-driven supersonic plasma jets 
in a collisionless regime for counter-streaming ions, in contrast to our recent 
oblique-jet-merging results that were in a much more collisional regime. 
\cite{merritt13prl,merritt14pop}   In this Letter, we identify collisionless plasma 
interpenetration transitioning to collisional stagnation between merging 
supersonic plasmas in an initially collisionless counter-streaming ion regime.  
Our diagnostic measurements, which span the jet-interaction region, are 
spatially and temporally well-resolved, allowing us to attribute the 
transition to an experimentally inferred rising mean ionization $\bar{Z}$ that drastically reduces the 
inter-jet ion--ion collisional mean free path, which scales as $\bar{Z}^{-4}$.  
The data are valuable for validating fundamental physics models of 
plasma collisionality\cite{jones96jcp} and ionization, especially in plasmas with 
complex equations of state (EOS). 

\begin{figure}
	\includegraphics[width=\linewidth]{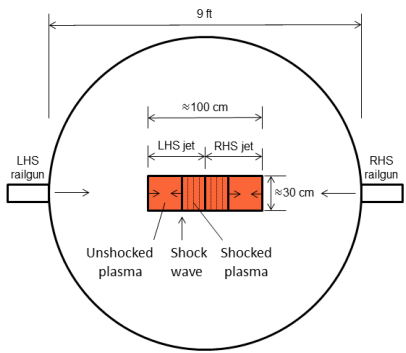}
	\caption{Cartoon of experimental setup.   
	\label{fig:geometry}}
	\end{figure}

Experiments are performed on the Plasma Liner Experiment.\cite{hsu12pop, 
hsu12ieee,hsu14jpp}  Pulsed-power-driven plasma 
railguns\cite{witherspoon11baps} launch jets from two directly opposed ports 
on a spherical vacuum chamber (Fig.~\ref{fig:geometry}). Each jet travels 
$\approx1.1$~m before they interact near chamber center ($z=0$~cm).  The 
working gas in the experiments presented here is argon.  The difference 
between chamber pressure rise after a plasma shot and after injection of 
only neutral gas suggests the possibility of significant impurity levels in the plasma jet.\cite{merritt13prl,merritt14pop}
Supported by observations of oxygen and aluminum 
spectral lines, we can reasonably deduce that impurities come from the railgun 
insulator material (zirconium-toughened alumina), and so we estimate
relative impurity percentages based on their relative abundance in the insulator.
An impurity sensitivity analysis comparing bounding cases of 60\% impurities, based on the chamber pressure difference, and 10\% 
impurities, chosen due to the appearance of impurity spectral lines at all times of interest, shows that 
our collisionality-based physics conclusions are independent of the impurity 
percentage assumed within these bounds (see Table~\ref{lengths}).  We also consider mixtures with small amounts of carbon and hydrogen, with no significant difference in results.  The analysis assuming a 40\% argon, 60\% impurities mixture provides the most conservative collision lengths (i.e., shortest) and so will be used here.
At the time of interaction the jets 
have ion density $n_i\sim10^{14}$~cm$^{-3}$, electron temperature 
$T_e\approx1$--$3$~eV, and relative velocity $v_{rel}\approx90$~km/s, with 
radius $\approx15$~cm and length $\approx50$~cm.  We expect the magnetic field 
to be $\sim1$~mT, based on extrapolation of the resistively decaying field 
measured along the railgun nozzle;\cite{merritt14pop} the ratio of magnetic 
field energy density to kinetic energy density is $\sim10^{-4}$ and so we treat 
the interaction as unmagnetized. The intra-jet thermal collisionality is high, 
but the relevant collision length for determining the nature of the interaction 
at initial jet merging is the inter-jet ion collision length, which we show is 
long.

\begin{figure*}
	\includegraphics[width=\linewidth]{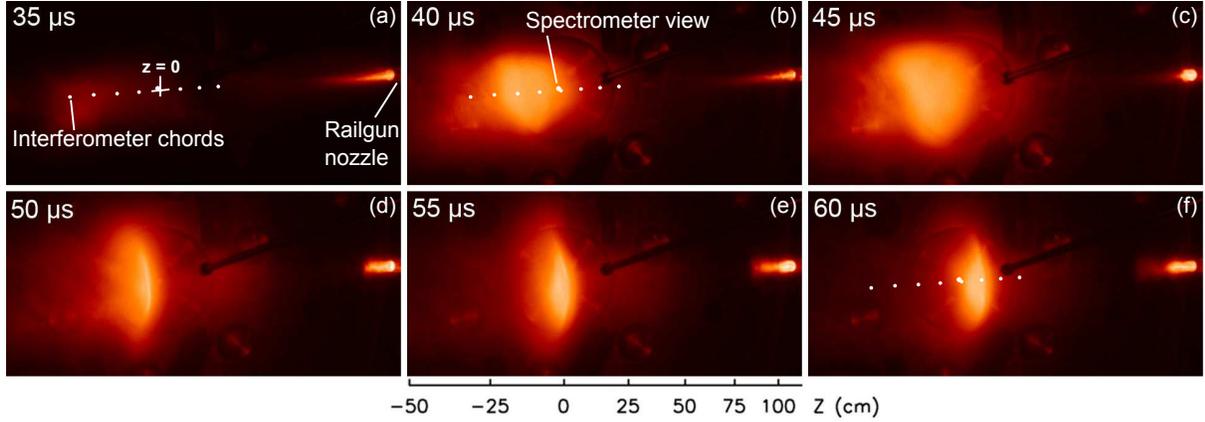}
	\caption{False-color fast-camera images (shots \#1834, 1833, 1836, 1837, 
	1838, 1845).  Images are 12 bit and have been logarithmically scaled.  
	Interferometer and spectrometer chord positions are indicated in (a), (b), 
	and (f) for comparison with Figs.~\ref{fig:int-time-series}(b), (g), and 
	(l), below.  Lineouts in Figs.~\ref{fig:lineout} and \ref{fig:stagnation} are taken from approximately along the line of the interferometer chords indicated in (b) and (f), respectively.\label{fig:time-series}}
	\end{figure*}

Diagnostics in the interaction region include a DiCam fast-framing (gate 20~ns) 
camera with a field of view reaching from one railgun nozzle to past chamber 
center ($\approx150$~cm), a SpectraPro survey spectrometer with a view diameter 
of $\approx1.5$~cm near $z=0$~cm, and an eight-chord laser interferometer\cite{merritt12rsia,merritt12rsib} with lines-of-sight spanning the 
interaction region at 7.5~cm intervals, from $z=-30$~cm to $z=22.5$~cm 
(Fig.~\ref{fig:time-series}).  The interferometer phase shift $\Delta\Phi$ is 
a line-integrated value sensitive to both free electrons and electrons bound in ions and neutrals; 
to determine $n_i$---here $\bar{Z}\geq 1$ and so $n_i\equiv n_e/\bar{Z}$---requires 
knowledge of $\bar{Z}$ and path length $\ell$, which is estimated from the 
full-width-half-maximum (FWHM) in a fast-camera image lineout at the spectrometer 
chord position divided by a factor of $\cos(30^{\circ})$ to account for the 
angle between interferometer line-of-sight and jet axis.  Spectrometer data and 
non-local-thermodynamic-equilibrium PrismSPECT\cite{macfarlane03, prism}
calculations together give a lower bound on peak $T_e$ and $\bar{Z}$ based on 
the appearance or absence of spectral lines.  Then $n_i$ is calculated using 
$\Delta\Phi=C_e (\bar{Z}-Err)\int n_{tot} d\ell$, where $n_{tot}=n_n+n_i$ is the total 
ion-plus-neutral density ($\approx n_i$ in all cases here), 
$C_e=\lambda e^2/4\pi\epsilon_{\text{0}}m_e c^2=1.58\times10^{-17}$~cm$^2$ is 
the phase sensitivity to electrons ($\lambda=561$~nm is the laser wavelength),
and $Err=0.08$ represents an upper bound on the phase sensitivity to ions.\cite{merritt14pop}  
PrismSPECT calculations are density 
dependent, and so the process is iterated until self-consistent $n_i$, $T_e$, 
and $\bar{Z}$ are reached.\cite{hsu12pop,merritt13prl}  

\begin{figure*}
	\includegraphics[width=\linewidth]{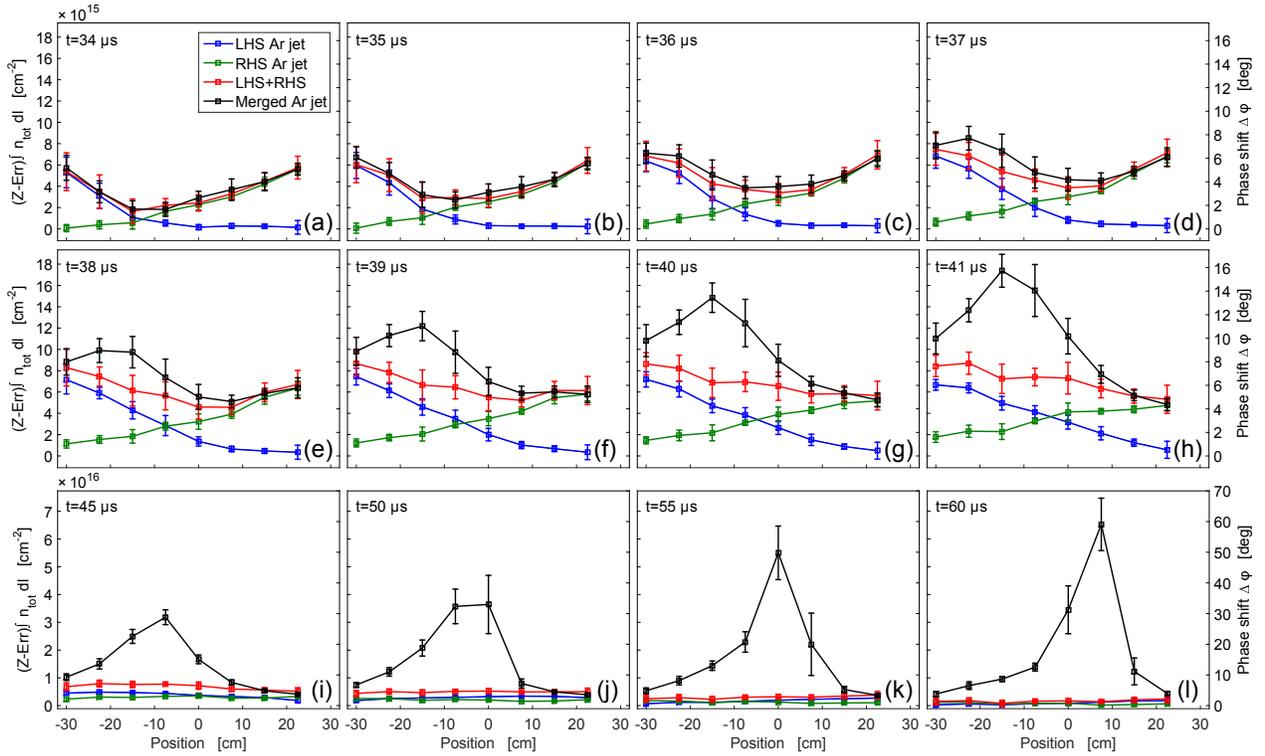}
	\caption{Time-resolved interferometer measurements comparing single-jet 
	(shots \#1846-1860)	and	merged-jet (shots \#1832-1845) experiments.  Comparing 
	average merged-jet traces (black) with the sum (red) of average single-jet traces (blue and green) shows 
	that the jets interpenetrate for $t=34$--$37$~$\mu$s (row 1), ionization increases the 
	merged-jet phase shift for $t=38$--$41$~$\mu$s (row 2), and the plasma has stagnated by 
	$t=60$ $\mu$s (row 3).  Error bars indicate standard deviation.  Note 
	change in vertical scale and inter-plot time step between rows 2 and 3.
	\label{fig:int-time-series}}
	\end{figure*}

Fast-camera images (Fig.~\ref{fig:time-series}) give an overview of the two-jet interaction 
from diffuse emission at $t=35$~$\mu$s to a 
bright, well-defined structure at $t=60$~$\mu$s. Interferometer measurements 
(Fig.~\ref{fig:int-time-series}; chords are into-the-page in Fig.~\ref{fig:time-series}) are 
from three sets of experiments: left-hand-side (LHS) jet only, averaged over 7 shots; right-hand-side 
(RHS) jet only, averaged over 7 shots; and merged-jet experiments, averaged over 14 shots.  
Shot-to-shot reproducibility is high;\cite{hsu12pop} error bars indicating standard deviation are small except for late times when they are increased due to slight shifts in position of steep gradients.  
Comparing these data sets 
(Fig.~\ref{fig:int-time-series}) allows us to compare the merged-jet case with 
the superposition of the two individual jets expected for simple 
interpenetration.  

Inter-jet collision lengths are calculated for ion--ion, ion--electron, and 
electron--electron collisions, taking all ion species into account; 
PrismSPECT calculations indicate that neutral density is $<0.3\%$ at all times presented here and so we are 
justified in neglecting neutrals.  Electron--electron collision length is 
$\ell^{e-e}=v_{th,e}/\nu_e$, using thermal collision frequency $\nu_e$ because
the electron thermal velocity $v_{th,e}\gg v_{rel}$.  We calculate both slowing and perpendicular 
collision length scales for ion--ion and ion--electron interactions, using 
$\ell=v_{rel}/\nu$ for all cases---except ion--ion slowing length scale 
$\ell^{i-i'}_s=v_{rel}/4\nu_s$\cite{messer13pop}---where the slowing frequency 
$\nu_{s}$ and perpendicular collision frequency $\nu_{\perp}$ are calculated in 
the slow limit for ion--electron and the fast limit for ion--ion.\cite{remark}  In this parameter range 
$\ell_{\perp} > \ell_s$, and so we present only the latter.  
Electron--electron collision length is 
$\ell^{e-e}=v_{th,e}/\nu_e$, using thermal collision frequency $\nu_e$ because
the electron thermal velocity $v_{th,e}\gg v_{rel}$. 
The total inter-jet ion--ion collision length for an ion species, taking 
interspecies collisions into account, is calculated by summing the collision 
frequencies for each collision type, e.g., for argon: 
$\ell_{s}^{\text{Ar}-i'}=
v_{rel}/\left[ 4 \,(\nu_{s}^{\text{Ar}-\text{Ar}}+
\nu_{s}^{\text{Ar}-\text{O}}+
\nu_{s}^{\text{Ar}-\text{Al}})\right]$.

\begin{table}
	\caption{\label{lengths}Experimentally inferred plasma parameters and 
	calculated collision lengths for the 40\% argon mixture (90\% argon mixture 
	values, where different, are shown in parentheses).  Collisionality 
	estimates are qualitatively unchanged for the two mixture assumptions.  
	Inclusion of 1\% each carbon and hydrogen in both the 40\% and 90\% argon mixtures, and inclusion of 5\% each carbon and hydrogen in the 40\% argon 
	mixture, leaves $n$, $T_e$, and $\bar{Z}$ 
	unchanged.}
	\begin{ruledtabular}
	\begin{tabular}{c c c c }
    	 & & t=35 $\mu$s & t=40 $\mu$s \\
    	\hline
    	$n_i$ ($10^{14}$~cm$^{-3}$) & & 1.5	(1.9) & 2.5 (2.4)		\\
    	$T_{e}$ (eV)			  & & 2.3 (1.7)	& 2.8	  \\
    	$\bar{Z}$				  & & 1.2 (1.0)	& 1.7 (1.8)	 \\
    	$Z_i$				& Ar	& 1.2	 (1.0) & 1.8 	\\
    						& O		& 1.0 		& 1.2	  \\
    						& Al	& 1.6	 (1.1) & 2.6  	\\					
    	\hline
		$\ell^{e-e}$ (cm) & 		& 0.054 (0.030)	& 0.035	(0.034)\\   	
 		\hline
     	$\ell_s^{i-e}$ (cm) &  Ar 	& 400 (360)	& 110		\\
     						&  O 	& 220 (140)	& 92 (90)	 	\\
     						&  Al 	& 160 (200)	& 37 (36)	 	\\	
     	\hline
     	$\ell_s^{i-i'}$ (cm) &  Ar 	& 1000 (2000)	& 140 (170)		\\
     						 &  O 	& 390 (500)	& 81	(92)	\\
     						 &  Al 	& 350 (950)	& 41	(48)	\\	
		\hline
		Interaction width (cm) & & 21--30 & 28--30	
	\end{tabular}
	\end{ruledtabular}
	\end{table}

In the case of ideal simple interpenetration, the interferometer trace in the two-jet
experiment will be the sum of the two single-jet experiment traces.
Figure \ref{fig:int-time-series} shows that the jet interaction is close to 
ideal simple interpenetration at early time, $t=34$--$37$~$\mu$s: $\Delta\Phi$ 
in two-jet experiments is nearly identical to the sum of single-jet experiment 
$\Delta\Phi$ at all $z$ positions for $t=34$--$36$~$\mu$s and within error bars 
for $t=37$~$\mu$s (Fig.~\ref{fig:int-time-series}). The merged-jet 
$\Delta\Phi=3.1^{\circ}$ at $z=0$~cm, $t=35$~$\mu$s 
(Fig.~\ref{fig:int-time-series}). Using $\bar{Z}=1.2$ inferred from spectroscopic 
measurements and PrismSPECT calculations, and $\ell=20$~cm, this $\Delta\Phi$ 
corresponds to $n_i=1.5\times10^{14}$~cm$^{-3}$.  Single-jet 
interferometer traces (Fig.~\ref{fig:int-time-series}) indicate that both jets contribute to the total merged 
$\Delta\Phi$, making this an upper bound on $n_i$ of each of the 
individual interpenetrating plasma jets.  We determine relative velocity of the two merging jets 
from the time-resolved single-jet interferometer traces:\cite{hsu12pop} a linear fit to the arrival time of 
$\Delta\Phi_{peak}$ at each chord gives a velocity of 41~km/s ($R^2=0.89$) for the LHS jet 
and 49~km/s ($R^2=0.48$) for the RHS jet.  This velocity represents the jet bulk; the 
diffuse leading edge interacting here is moving at a higher velocity due to jet 
expansion, so the quoted velocities are a lower bound at $t=35$ $\mu$s, giving a conservative underestimate of collision lengths.
	
\begin{figure}
	\includegraphics[width=\linewidth]{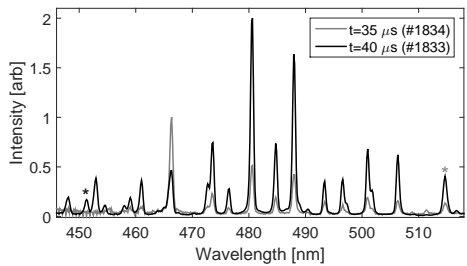}
	\caption{\label{fig:spec}Spectroscopic measurements for $t=35$~$\mu$s (shot \#1834) and $t=40$~$\mu$s (shot \#1833), near $z=0$~cm [see Fig.~\ref{fig:time-series}(a) \& (b) for spectrometer chord position].  Lines used to determine $T_e$, 514.7 nm (Ar II) for the $t=35$~$\mu$s case and 451.4 nm (Al III) for the $t=40$~$\mu$s case, are indicated with an asterisk.}
	\end{figure}

The appearance of emission at chamber center in two-jet experiments, but not in 
single-jet experiments, suggests that the jets do interact, if minimally, as 
they interpenetrate at early time.  This is confirmed by spectroscopic 
measurements at $z=0$~cm, $t=35$~$\mu$s, which give $T_{e}=2.3$~eV [based on the appearance of the Ar II line at 514.7 nm, (Fig.~\ref{fig:spec})],   
greater than the $T_{e}=1.9$~eV measured at the railgun nozzle.  The 
experimentally inferred $\bar{Z}=1.2$ is unchanged from nozzle to interaction 
region, so the only measurable interaction effect as the jets interpenetrate is 
a slight temperature increase.

The observation of jet interpenetration suggests that collisionality between ions of one jet and ions and electrons of the opposing jet is low.  We verify this by calculating inter-jet collision lengths and comparing them to the scale size of the experiment.
Using $v_{rel}=90$~km/s, $n_i=1.5\times10^{14}$~cm$^{-3}$, $T_{e}=2.3$~eV, and 
$\bar{Z}=1.2$, we calculate inter-jet collision lengths, presented in Table 
\ref{lengths}.  All ion collision lengths for $t=35$~$\mu$s are significantly 
longer than the length scale of the experiment, which we estimate as $\approx21$ cm from a fast-camera image lineout or $\approx30$ cm based on interferometer values. This is consistent with the 
observation that the jets interpenetrate.
	
	\begin{figure}
	\includegraphics[width=\linewidth]{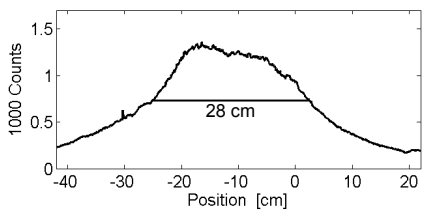}
	\caption{\label{fig:lineout}Fast-camera image [Fig.~\ref{fig:time-series}(b)] lineout vs.\ $z$ through $z=0$~cm interferometer chord for $t=40$~$\mu$s (shot \#1833).  Horizontal line indicates FWHM=28~cm.}
	\end{figure}

As the higher-density bulk of the jet arrives at chamber center (leading edges 
are interpenetrating at $t=35$~$\mu$s; $\Delta\Phi_{peak}$ reaches chamber 
center at $t\approx42$ and $46$~$\mu$s for the LHS and RHS jet, respectively), 
the collision scale lengths drop and $\bar{Z}$ increases.  At $t=40$~$\mu$s the 
two-jet $\Delta\Phi$ is greater than the sum of single-jet $\Delta\Phi$ in the 
region from $z=-22.5$ to $-7.5$~cm (Fig.~\ref{fig:int-time-series}).
 
An increase in merged-jet $\Delta\Phi$ over the sum of single-jet $\Delta\Phi$ 
indicates that the jets are no longer simply interpenetrating.
The merged-jet phase shift at $z=0$~cm, $t=40$~$\mu$s is 
$\Delta\Phi=8.1^{\circ}$, for which our iterative process 
(using $\ell=22$~cm) gives $n_i=2.5\times10^{14}$~cm$^{-3}$, 
$T_{e}=2.8$~eV [based on the appearance of the Al III line at 451.4 nm, (Fig.~\ref{fig:spec})], and $\bar{Z}=1.7$.  Again, because the jets have 
interpenetrated, this bounds the $n_i$ of the individual jets.
Table ~\ref{lengths} lists collision scale lengths calculated with these values.  The 
increase in $\Delta\Phi$ in the merged-jet case over the 
simple-interpenetration case is consistent with a $\bar{Z}$ increase rather 
than an $n_i$ increase at $t=40$~$\mu$s.  The inferred $\bar{Z}=1.7$--$1.8$ is a 
factor of $\approx 1.4$--$1.8$ greater than the $\bar{Z}=1.0$--$1.2$ for 
interpenetrating jets at $t=35$~$\mu$s.\cite{ionization-note}  The ratio of the $\Delta\Phi$ in the 
merged-jet case to $\Delta\Phi$ for the sum of single jets at $z=0$~cm, 
$t=40$~$\mu$s is $8.1^{\circ}/5.9^{\circ}\approx1.4$; thus, increased 
ionization is sufficient to account for the increase in $\Delta\Phi$ between 
the two cases.  

Interferometer measurements at $t=40$~$\mu$s indicate that the width of the 
$\Delta\Phi$ increase is 15--30~cm (the difference between merged-jet and 
single-jet $\Delta\Phi$ exceeds error bars at $z=-22.5,-15,-7.5$~cm but not at 
$z=-30, 0$~cm).  The region of increased emission in fast camera images aligns 
with the increased $\Delta\Phi$ in interferometer measurements [compare 
Figs.~\ref{fig:time-series}(b) and \ref{fig:int-time-series}(g)].  A 
horizontal lineout of the fast-camera image shows that the region of increased 
emission has a FWHM of 28~cm (Fig.~\ref{fig:lineout}).  Both of these estimates 
are of the same order as $\ell_s^{\text{Al}-i'}\approx\ell_s^{\text{Al}-e}\approx40$~cm at
$t=40$~$\mu$s. 
Because the slowing lengths have now dropped to the order 
of the interaction width, we expect to see plasma stagnation as the interaction 
progresses.

\begin{figure}
	\includegraphics[width=\linewidth]{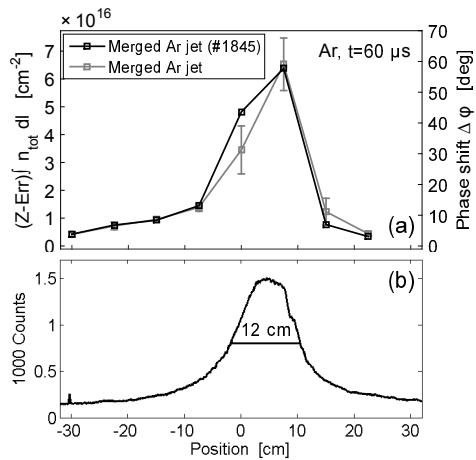}
	\caption{\label{fig:stagnation}(a) Interferometer measurements for 
	$t=60$~$\mu$s:  merged-jet shot \#1845 (black) and average of merged-jet shots (gray). Error bars indicate standard deviation.  Structure is consistent with shock formation, with the peak in the black trace corresponding to postulated post-shock plasma, bounded on both sides by postulated shocks (the drop-off between chords at $z=0$ and $-7.5$~cm and chords at $z=7.5$ and 15~cm). (b) Fast-camera image lineout [Fig.~\ref{fig:time-series}(f)] vs.\ $z$ through
the	$z=0$~cm interferometer chord for $t=60$~$\mu$s (shot \#1845).  Increased emission aligns with peak, suggesting post-shock plasma in (a).  Horizontal line indicates FWHM=12~cm.}
	\end{figure}

Plasma stagnation leads to formation of a large, pronounced peak in the interferometer 
trace, with $\Delta\Phi_{peak}=59.1^{\circ}$, by $t=60$~$\mu$s.  Because shot-to-shot 
variation in $\Delta\Phi$ increases at later times, the interferometer data for 
the individual shot shown in the fast-camera image in 
Fig.~\ref{fig:time-series}(f) (\#1845) is plotted along with the 14-shot-averaged 
data in 
Fig.~\ref{fig:stagnation}(a).  For the individual trace, a pronounced peak 
spanning two interferometer chords drops off to $1/8$--$1/3$ the peak value on 
either side ($\Delta\Phi=43.6^{\circ}$ and $57.7^{\circ}$ to 
$\Delta\Phi=13.1^{\circ}$ and $7.0^{\circ}$).  This peak aligns with the 
region of increased emission in Fig.~\ref{fig:int-time-series}(f); a lineout of 
the image [Fig.~\ref{fig:stagnation}(b)] shows that the increased emission has a 
FWHM of 12~cm and is centered between the two high-$\Delta\Phi$ chord 
positions.  The spectrometer view is near $z=0$~cm at the 
peak edge and measures $\Delta\Phi=43.6^{\circ}$; with $\ell=29$~cm this corresponds 
to $n_i=1.3\times10^{15}$~cm$^{-3}$, $T_{e}=2.2$~eV, and $\bar{Z}=1.4$.  
A jet velocity of 45~km/s
corresponds to Mach number $M\approx 8.5$ (calculated with $t=40$~$\mu$s plasma values), implying minimal expansion of the plasma normal to the jet velocity.  Because $\ell=29$~cm may be an underestimate, we also evaluate plasma conditions assuming $\ell=44$~cm (estimated from the full-width-10\%-maximum), obtaining $n_i=6.6\times10^{14}$~cm$^{-3}$, $T_{e}=2.4$~eV, and $\bar{Z}=1.5$.

The stagnated plasma has led to structure consistent with collisional shocks by 
$t=60$~$\mu$s, as inferred from the observed $n_i$-transition scale of 
$\leq 7.5$~cm [Fig.~\ref{fig:stagnation}(a)] being comparable to
the predicted shock thickness, which is of order the post-shock thermal ion 
mean-free-path $\lambda_{mfp,i}$.\cite{jaffrin64pof}  We use a one-dimensional shock model, which is
justified based on our observations of minimal expansion normal to the jet velocity, to
estimate post-shock ion temperature.  
Assuming that $T_e$ does not change across the shock,\cite{jaffrin64pof} this predicts post-shock $T_i=61$~eV for both cases calculated above, 
and shock width $\lambda_{mfp,i}\approx2$--$3$~cm, consistent with the observed 
scale of $\leq7.5$~cm.
	
In summary, we have presented a concrete example of colliding supersonic 
plasmas in an initially collisionless regime for counter-streaming
ions, transitioning from collisionless interpenetration to collisional 
stagnation owing to dynamically rising $\bar{Z}$. In the interpenetration 
stage, inter-jet collision lengths are much greater than the experimental 
length scale, consistent with simple interpenetration seen in interferometer 
measurements. In the increasing-$\bar{Z}$ phase,
the inter-jet $\ell_s^{i-e}$ and $\ell_s^{i-i^\prime}$ are on the same order as
the emission FWHM
and width of the interferometer $\Delta \Phi$ peak.
Finally, in the stagnation stage, $\ell_s^{i-i^\prime}$ values decrease to the 
interaction region width, leading to structure consistent with collisional shock formation. These 
measurements provide an opportunity to validate fundamental physics models 
used to calculate plasma collisionality\cite{jones96jcp} and ionization in 
plasmas with complex EOS.

\begin{acknowledgments} 
This work was supported by the LANL LDRD Program under DOE contract 
no.\ DE-AC52-06NA25396.  We acknowledge J.~P. Dunn and E.~C. Merritt for technical support, and C.~S. Adams and I. Golovkin for technical support and useful discussions.
\end{acknowledgments}

%


\begin{thebibliography}{32}
\bibitem [{ion()}]{ionization-note}%
  \BibitemOpen
  \href@noop {} {}\bibinfo {note} {The observed rise in $\bar{Z}$ on an
  $\approx 5$-$\mu$s time scale cannot be readily explained by simple
  electron-impact (e.g., based on rate-coefficient data\cite{chung05}) nor
  ion-impact\cite{hasted72} ionization. Further work beyond the scope of this
  paper is needed to detail the probable ionization mechanisms.}\BibitemShut
  {Stop}%
\end{thebibliography}
\end{document}